\documentclass{article}[11pt]

\usepackage{amsfonts,amsmath,amssymb,graphicx,epsfig,color,setspace} 

\newcommand{\be}{\begin{equation}}
\newcommand{\ee}{\end{equation}}
\newcommand{\ba}{\begin{eqnarray}}
\newcommand{\ea}{\end{eqnarray}}

\begin{document}

\onehalfspacing

\thispagestyle{empty}

\begin{center}

{\Large \bf The Noise of Gravitons}

\bigskip
\bigskip
\bigskip

{\large \sc Maulik~Parikh$^{*\times}$\footnote{{\tt maulik.parikh@asu.edu}}, Frank Wilczek$^{*\dagger}$\footnote{\tt frank.wilczek@asu.edu}, and George Zahariade$^\times$\footnote{\tt george.zahariade@asu.edu}}

\bigskip
\bigskip

{\em $^*$Department of Physics, Arizona State University, Tempe, Arizona 85287, USA} \\
{\em $^{\times}$Beyond Center for Fundamental Concepts in Science, Arizona State University, Tempe, Arizona 85287, USA} \\
{\em $^\dagger$Department of Physics, Stockholm University, Stockholm SE-106 91, Sweden} \\
{\em $^\dagger$Center for Theoretical Physics, Massachusetts Institute of Technology, Cambridge, Massachusetts 02139, USA} \\
{\em $^\dagger$Wilczek Quantum Center, Department of Physics and Astronomy, Shanghai Jiao Tong University, Shanghai 200240, China}

\bigskip
\bigskip
\bigskip

{\bf Abstract} \\
\noindent
\end{center}
We show that when the gravitational field is treated quantum-mechanically, it induces fluctuations -- noise -- in the lengths of the arms of gravitational wave detectors. The characteristics of the noise depend on the quantum state of the gravitational field, and can be calculated exactly in several interesting cases. For coherent states the noise is very small, but it can be greatly enhanced in thermal and (especially) squeezed states.  Detection of this fundamental noise would constitute direct evidence for the quantization of gravity and the existence of gravitons.  
\bigskip
\bigskip
\bigskip

\begin{center}
\noindent
{\em This essay was awarded First Prize in the 2020 Essay Competition of the Gravity Research Foundation.}
\end{center}

\newpage

\setcounter{page}{1}

\noindent
While there is much we don't know about the complete theory of quantum gravity, we do know that perturbation theory, when applied to Einstein gravity, contains a massless helicity-two particle: the graviton. Conversely, consistency in the quantum mechanics of a massless helicity-two particle leads one to Einstein gravity, as Feynman famously argued \cite{Feynman:1963ax,Weinberg:1964kqu,Deser:1969wk,Boulware:1974sr}. However, experimental support for the existence of gravitons remains weak. Clearly, it would be very desirable to identify empirical phenomena that could be attributed convincingly to the quantization of the gravitational field or, in other words, to the existence of gravitons.   

With this in mind, we have calculated the behavior of gravitational wave detectors \cite{Abbott:2016blz,Audley:2017drz}, in response to {\it quantized\/} states of the gravitational field  \cite{shortpaper,longpaper}. The quantum nature of the gravitational field manifests itself as a characteristic state-dependent noise. For coherent states the noise is tiny, as anticipated by Dyson \cite{Dyson:2013jra}, but in other kinds of states it can be significantly larger, and potentially detectable. 

We model the detector as two free-falling masses whose geodesic separation is being monitored. According to the geodesic deviation equation, the separation of the masses is sensitive to the Riemann tensor induced by gauge-invariant perturbations of the metric, including by incident gravitational waves. Let the geodesic separation be $\xi$. Then
\be
\ddot{\xi} = \frac{1}{2} \ddot{h} \xi \ ,
\label{classeom}
\ee
where $h$ is the metric perturbation, or strain. This familiar equation, known mathematically as a Hill equation, gives the tidal acceleration of $\xi$ in the presence of a gravitational wave. By solving Einstein's equations with different sources, one obtains signal templates for the strain $h$, which in turn feed the celebrated stretching and squeezing of $\xi$, the length of the detector arm.

The question we would like to ask now is: how does this equation change when the gravitational field is quantized? Or, put another way, what is the equation of motion for the geodesic deviation if the spacetime metric is actually a quantum field?

To answer this, we go back to basics. Our gravity+detector system is described by the Einstein-Hilbert action coupled to the actions of two free-falling masses, $M_0$ and $m_0$, with worldlines $X^\mu(t)$ and $Y^\mu(t)$:
\be 
S = \frac{1}{16 \pi G} \int d^4 x \sqrt{-g} R - M_0 \int dt \sqrt{-\dot{X}^2} - m_0 \int dt \sqrt{-\dot{Y}^2} \; .
\ee
If the two masses are close enough, the metric can be regarded as nearly flat: $g_{\mu \nu} = \eta_{\mu \nu} + h_{\mu \nu}$. We now expand the action to leading order and make some judicious gauge choices. We also decompose the perturbation, $h(t,\vec{x})$, into Fourier modes with amplitude $q(t)$ and frequency $\omega$. Focusing for now on a single mode (and with some additional simplifying assumptions, such as restricting to one polarization), the action reduces \cite{longpaper} further to 
\be
S_\omega = \int dt \left ( \frac{1}{2} m (\dot{q}^{ 2} - \omega^2  q^2) + \frac{1}{2} m_0  \dot{\xi}^2 - g \dot{q} \dot{\xi} \xi  \right ) \, .\label{Somega}
\ee 
This is the action for a gravitational field mode of energy $\hbar\omega$, with amplitude proportional to $q$, interacting with a free-falling mass $m_0$ whose geodesic separation (``arm length'') from a heavier  mass is given by $\xi$. The unphysical mass $m$, introduced for dimensional reasons, will play no role, and the coupling constant $g$ is proportional to $m_0$, in accordance with the equivalence principle. The action describes a simple harmonic oscillator coupled to a free particle via a cubic derivative interaction. It is ready for quantization.  

Before plunging in, let us anticipate our strategy. We have a harmonic oscillator (the gravitational field mode) in some initial state, $|\psi_\omega\rangle$. The mode can have a final state $|f \rangle$, which, after interaction with the detector, will generically be different from its initial state because the detector masses will typically both absorb and emit gravitons (through spontaneous as well as stimulated emission). However, we are not interested in the final state of the mode; indeed, the only quantity we can directly measure is the arm length $\xi$ itself. That is, we would like to integrate out the graviton mode as well as sum over its final states. Thus, the most general thing we can calculate is the transition probability for $\xi$ to go from state $|\phi_A\rangle$ to state $|\phi_B\rangle$ with an interaction that takes place between $t = 0$ and $t = T$:
\be
P_{\psi_\omega}(\phi_A \to \phi_B)  =  \sum_{|f\rangle} |\langle f, \phi_B | \hat{U}(T,0) | \psi_\omega, \phi_A \rangle |^2\,.
\ee
Here $\hat{U}$ is the unitary time-evolution operator associated with the Hamiltonian obtained from~\eqref{Somega}, and our notation for tensor product states of the joint Hilbert space is $|a,b\rangle \equiv |a\rangle\otimes|b\rangle$.

The evaluation of this transition probability is a calculation in ordinary quantum mechanics. Due to the relatively simple form of the Lagrangian,~\eqref{Somega}, which is quadratic in $q$, it can be evaluated exactly \cite{longpaper}, without recourse to perturbation theory, in either the canonical or path-integral approach. The calculation is not entirely straightforward, however, because there are several subtleties along the way, in particular the derivative coupling and the finite-time interaction. When the dust settles, we obtain
\be
P_{\psi_\omega}(\phi_A \to \phi_B) \sim \int {\cal D}\xi {\cal D}\xi' e^{\frac{i}{\hbar} \int_{0}^{T} dt \frac{1}{2} m_0 (\dot{\xi}^2 - \dot{\xi}^{'2})} F_{\psi_\omega}[\xi,\xi']\ .
\label{nextprob}
\ee
This is a double path integral because it describes a probability, rather than an amplitude. In the exponent is the free action of the particle. The gravitational field mode has been integrated out and the entirety of its effect on the arm length is encapsulated in $F_{\psi_\omega}[\xi,\xi']$, known as the Feynman-Vernon influence functional~\cite{Feynman:1963fq}. The aptly named influence functional captures the effect, or influence, of one quantum system on another. In our context, the influence functional tells us about the effect of the quantized gravitational field mode, in an initial state $|\psi_\omega\rangle$, on the physics of the detector arm length. We can derive a compact analytic expression for $F_{\psi_\omega}[\xi,\xi']$; crucially, we are able to evaluate it explicitly for several interesting classes of states \cite{longpaper}.

Now we can tackle the general problem of a continuum of modes -- a quantum field -- interacting with the detector. The quantum state of the gravitational field $|\Psi\rangle$ can be written as a tensor product of the Hilbert states of the individual graviton modes: $|\Psi \rangle = \bigotimes_{\vec{k}}|\psi_{\omega(\vec{k})}\rangle$.
Since the action for the field involves a sum over modes, the field-theoretic influence functional becomes a product of the quantum-mechanical mode influence functionals. For several important classes of states (the vacuum, coherent states, thermal states, squeezed states), we are able to evaluate the full influence functional. This typically contains a term of the form 
\be
F_{\Psi}[\xi,\xi'] \sim e^{\frac{1}{2 \hbar^2} \int A_\Psi (X - X')^2}
\ee
where $X, X'$ are some known functions of $\xi, \xi'$.

Next comes an ingenious trick. Following Feynman and Vernon, we can express the influence functional in a remarkably suggestive form. We insert the identity
\be
e^{\frac{1}{2 \hbar^2} \int A_\Psi (X - X')^2} \sim \int {\cal D} N_\Psi e^{- \frac{1}{2} \int A^{-1}_\Psi N_\Psi^2 + \frac{i}{\hbar} \int N_\Psi (X - X')}
\ee
into our transition probability and (disregarding some technicalities \cite{longpaper} for brevity) find roughly that
\be
P_{\Psi}(\phi_A \to \phi_B) \sim
\int {\cal D} N_\Psi e^{- \frac{1}{2} \int A^{-1}_\Psi N_\Psi^2} \left | \int {\cal D}\xi e^{\frac{i}{\hbar} \int_{0}^{T} dt \left ( \frac{1}{2} m_0 \dot{\xi}^2 + \frac{1}{4} m_0 (\ddot{h} + \ddot{N}_\Psi) \xi^2 \right )} \right |^{2} \ .
\ee
This expression tells us something remarkable. It says that the detector arm length is subject to an {\em additional} fluctuation. This extra function $N_\Psi(t)$ can be viewed as Gaussian noise. The statistical properties of the noise stem from aspects of $A_\Psi$, its auto-correlation function. Thus we see that the upshot of integrating out a quantum field is to couple the remaining degree of freedom, $\xi$, to stochastic noise \cite{Hu:1999mm,Calzetta:1993qe}. {\em It is the noise produced by the quantization of the gravitational field: the noise of gravitons.}

Finally, we can derive an effective, quantum-corrected equation of motion for the arm length $\xi$ by taking a saddle point of the $\xi$ path integral. The result, when done carefully \cite{longpaper}, is
\be
\ddot{\xi} = \frac{1}{2}\left(\ddot{h}+\ddot{N}_\Psi-\frac{m_0G}{c^5}\frac{d^5}{dt^5}\xi^2 \right)\xi \ .
\label{langevineq}
\ee
This striking equation is the quantum generalization of~\eqref{classeom}. It is essentially the geodesic deviation equation in the presence of a {\em quantized} gravitational field.
Within the parentheses are three terms that source the tidal acceleration $\ddot{\xi}$. The first term is the classical gravitational perturbation, already encountered in \eqref{classeom}. The last term is a gravitational radiation reaction term, the counterpart of the three-derivative Abraham-Lorentz acceleration in electromagnetism. The pathologies that ensue when radiation reaction equations are taken literally have been the subject of much confusion, and it has long been anticipated that quantum effects will somehow remedy the situation. Here we see that such equations arise as approximations to path integrals that are free of pathologies. 

Most interestingly, \eqref{langevineq} contains a state-dependent quantum noise, $N_\Psi(t)$, as a source. The presence of this term means that this is now a {\em stochastic} differential equation. That is intuitively appealing: it conforms to the expectation that a quantum field will induce random fluctuations in any classical degree of freedom it interacts with. This randomness has the effect of altering the dynamics of the classical degree of freedom so that it is necessarily described by a stochastic -- rather than a deterministic -- equation of motion. 

The properties of the noise -- its amplitude, power spectrum, etc -- are calculable and depend on the state. We find that for the vacuum state or a coherent state, the fluctuations in the arm length are extremely small and almost certainly undetectable, as foreseen by Dyson. But for thermal states -- such as from cosmology or evaporating black holes -- the noise is significantly enhanced.  Most favorably, if the gravitational field is in a squeezed state, as predicted by some inflationary models, the fluctuations in the arm length can be enhanced by an exponential of the squeezing parameter, and are potentially detectable.  

The study of noise has historically played an important role in several major developments in physics. It was unexplained noise in a radio receiver that led Penzias and Wilson to discover the cosmic microwave background. It was noise that supplied early evidence for the existence of molecules (through Brownian motion), and for the existence of fractionally-charged quasiparticles (through shot noise). It is possible, likewise, that the existence of gravitons will first be revealed through noise.
\newpage

\noindent {\it Acknowledgments.} We thank Paul Davies, Bei-Lok Hu, Phil Mauskopf, and Tanmay Vachaspati for conversations. During the course of this work, 
MP and GZ were supported in part by John Templeton Foundation grant 60253. GZ also acknowledges support from the 
Foundational Questions Institute and Moogsoft. FW is supported in part by the U.S. Department of Energy under grant DE-SC0012567, 
by the European Research Council under grant 742104, and by the Swedish Research Council under contract 335-2014-7424.

\end{document}